\documentclass[twocolumn]{aastex63}
\usepackage{bm}
\usepackage{url}
\usepackage{amsmath}
\usepackage{color}
\usepackage{graphicx}

\received{}
\revised{}
\accepted{}
\submitjournal{ApJ}

\shorttitle{CW Tau}
\shortauthors{Ueda et al.}

\begin{document}

\title{
Massive compact dust disk with a gap around CW Tau revealed by ALMA multi-band observations
}

\correspondingauthor{Takahiro Ueda}
\email{takahiro.ueda@nao.ac.jp}

\author[0000-0003-4902-222X]{Takahiro Ueda}
\affil{National Astronomical Observatory of Japan, Osawa 2-21-1, Mitaka, Tokyo 181-8588, Japan}

\author[0000-0003-4562-4119]{Akimasa Kataoka}
\affil{National Astronomical Observatory of Japan, Osawa 2-21-1, Mitaka, Tokyo 181-8588, Japan}

\author[0000-0002-6034-2892]{Takashi Tsukagoshi}
\affil{National Astronomical Observatory of Japan, Osawa 2-21-1, Mitaka, Tokyo 181-8588, Japan}



\begin{abstract}
Compact protoplanetary disks with a radius of $\lesssim$ 50 au are common around young low-mass stars.
We report high resolution ALMA dust continuum observations toward a compact disk around CW Tau at Band 4 ($\lambda=2.2$ mm), 6 (1.3 mm), 7 (0.89 mm) and 8 (0.75 mm).
The SED shows the spectral slope of $2.0\pm0.24$ between 0.75 and 1.3 mm, while it is $3.7\pm0.29$ between 2.17 and 3.56 mm.
The steep slope between 2.17 and 3.56 mm is consistent with that of optically thin emission from small grains ($\lesssim$ 350 ${\rm \mu m}$).
We perform parametric fitting of the ALMA data to characterize the dust disk.
Interestingly, if the dust-to-gas mass ratio is 0.01, the Toomre's Q parameter reaches $\sim$ 1--3, suggesting that the CW Tau disk might be marginally gravitationally unstable.
The total dust mass is estimated as $\sim250M_{\oplus}$ for the maximum dust size of 140 ${\rm \mu m}$ that is inferred from the previous Band 7 polarimetric observation and at least $80M_{\oplus}$ even for larger grain sizes.
This result shows that the CW Tau disk is quite massive in spite of its smallness.
Furthermore, we clearly identify a gap structure located at $\sim20$ au, which might be induced by a giant planet.
In spite of these interesting characteristics, the CW Tau disk has {\it normal} disk luminosity, size and spectral index at ALMA Band 6, which could be a clue to the mass budget problem in Class II disks.
\end{abstract}


\section{Introduction}
The observations with Atacama Large Millimeter/submillimeter Array (ALMA) have revealed that axisymmetric gap and ring structures are common in the dust continuum emission of protoplanetary disks (e.g., \citealt{ALMA+15,Andrews+18,Long+18}), which invokes that giant planets might be forming within them (e.g., \citealt{Pinilla+12,Dipierro+15}).
Although many studies have focused on relatively large disks which are preferable as a laboratory to characterize the substructures, recent survey studies have shown that compact disks ($\lesssim$ 50 au) are more common in low-mass star-forming regions than larger ones ($\sim$ 70--90\%; \citealt{Barenfeld+17,Cieza+19,Long+19,Otter+21}). 

The Spectral Energy Distribution (SED) at mm wavelengths has been used for characterizing dust disks because the spectral slope $\alpha$ reflects the slope of the absorption opacity $\beta$ (and hence the dust size) in optically thin limit
 (e.g., \citealt{Beckwith+90,BS91,Testi+03,Rodmann+06,Isella+09,Ricci+10,Guilloteau+11}).
Given the sufficient angular resolution, the SED approach can be extended to the spatially-resolved characterization of disks (e.g., \citealt{Perez+12,Tazzari+16,Tsukagoshi+16,Carrasco+19,Ueda+20,Ueda+21,Macias+21,Sierra+21}).
One of the interesting findings of these works is that most of the targeted disks has a dust surface density massive enough to be optically thick at sub-mm wavelengths (e.g., ALMA Band 7).

Recent theoretical studies have shown that the brightness of optically thick disks can be reduced by scattering of dust thermal emission, which makes them look optically thin \citep{MN93,Liu+19,Zhu+19}.
When one tries to estimate dust properties from the observed intensity, the emergent intensity is often assumed as a function of three parameters; dust surface density, dust temperature, spectral slope of absorption opacity.
However, the scattering contributes to the intensity as an additional parameter, meaning that analysis with more than four wavelengths is essential to solve the degeneracy between these parameters.
Although compact disks are also supposed to have substructures potentially induced by forming-planets as larger population does (e.g., \citealt{Yamaguchi+21}), the multi-wavelength characterization toward compact disks is still sparse due to the limited resolution.

In this paper, we present an ALMA multi-band observations toward CW Tau.
CW Tau is a young T Tauri star located in Taurus star-forming region.
The stellar parallax is estimated as $7.60\pm0.04$ mas \citep{Gaia+21}, corresponding to the distance of 130.9--132.3 pc.
In the following, we adopt the value of 132 pc.
The spectral type is estimated as K3 \citep{Luhman+10} and the stellar luminosity is $\sim2.4L_{\odot}$ \citep{Andrews+13}, inferring a stellar age of $\sim2.2$ Myr \citep{Pietu+14}.
The CW Tau disk has been spatially resolved by ALMA Band 7 polarimetric observation with an angular resolution of 0$\farcs$24$\times$0$\farcs$14 corresponding to a spatial resolution of $\sim$ 26 au \citep{Bacciotti+18}.
The details of the observations and processes of data reduction and imaging are described in Section \ref{sec:obsdata}.
Section \ref{sec:results} describes the results of the observations.
We model the dust disk with a parametric fitting of the observed images in Section \ref{sec:diskmodelling}.
Discussion and conclusion are in Section \ref{sec:discussion} and \ref{sec:conclusion}, respectively.

\begin{figure*}[ht]
\begin{center}
\includegraphics[scale=0.47]{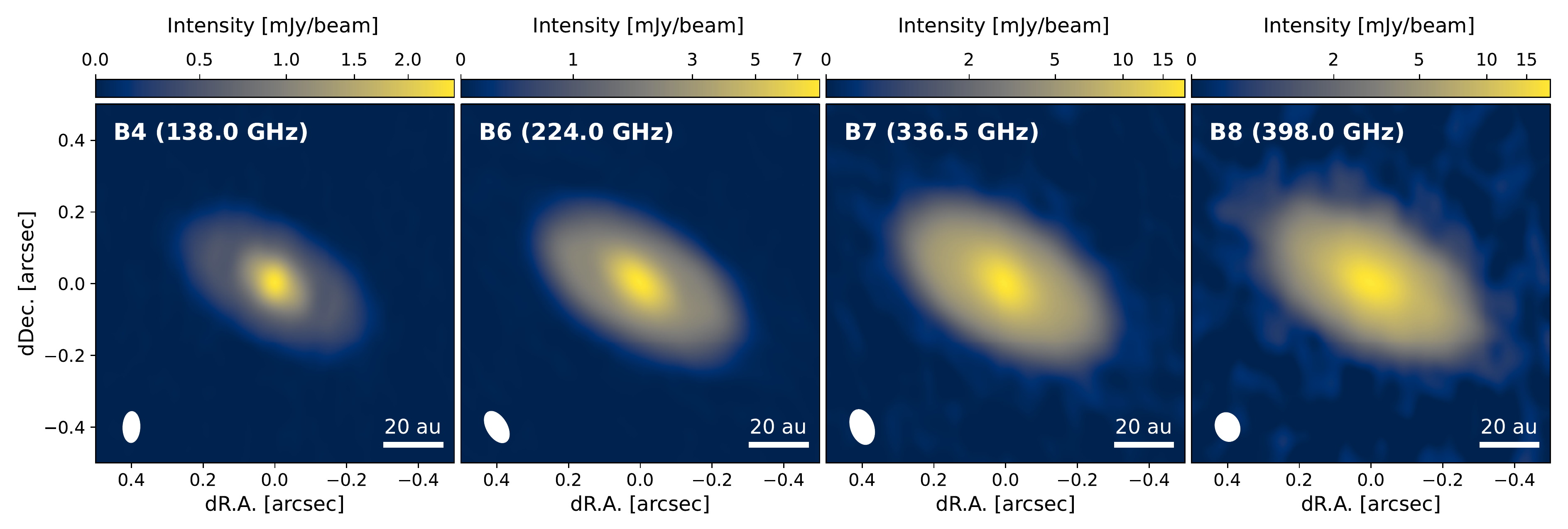}
\caption{
ALMA observations of the CW Tau disk at Band 4, 6, 7 and 8.
The synthesized beam size is denoted with a white filled ellipse at lower left in each panel.
An arcsinh stretch is applied to the color scale of each disk to increase the visibility of a gap in the disk.
}
\label{fig:images}
\end{center}
\end{figure*}

\section{Observations} \label{sec:obsdata}

\subsection{Observational data}
We observed dust continuum emission around CW Tau in the ALMA
Cycle 7 program 2019.1.01108.S (PI: T. Ueda). 
Observations were carried out in Band 4, 6, 7 and 8.
The spectral setup included four spectral windows with 2.0 GHz wide, centred at the standard ALMA frequencies: 138, 140, 150 and 152 GHz for Band 4; 224, 226, 240 and 242 GHz for Band 6; 336, 338, 348 and 350 GHz for Band 7; 398, 400, 410 and 412 GHz for Band 8.
The Band 4 data includes two execution blocks made on 2021 September 15 and 30.
The Band 6 data includes two execution blocks made on 2021 August 8 and 9.
The Band 7 and 8 observations have been done with one execution block on 2021 July 22 and 2021 July 15, respectively.
Total exposure time was 156.9, 170.5, 19.4 and 40.4 min at Band 4, 6, 7 and 8, respectively.
The phase calibrator was quasar J0403+2600 for Band 4 and 6, J0438+3004 for Band 7 and 8.
The bandpass calibrator was quasar J0510+1800 for Band 4 and 7, J0237+2848 for Band 6 and J0423-0120 for Band 8.
The details of the observation setup are summarized in Table \ref{table:0}.

\begin{table*}[ht]
  \begin{center}
  \caption{Observation setup}
  \label{table:0}
  \begin{tabular}{cccccc}
  \hline
 Band & central freqencies & date & on-source time & phase calibrator & bandpass calibrator\\
      &     (GHz)            &   &   (min)   &  &       \\
\hline\hline
   4   & 138, 140, 150, 152 & 15 Sep. 2021 & 42.23 & J0403+2600  & J0510+1800 \\
       &                    & 30 Sep. 2021 & 42.17 & J0403+2600  & J0510+1800 \\
   6   & 224, 226, 240, 242 & 8 Aug. 2021  & 47.43 & J0403+2600 & J0237+2848 \\
       &                    & 9 Aug. 2021  & 47.47 & J0403+2600 & J0237+2848 \\
   7   & 336, 338, 348, 350 & 22 Jul. 2021 & 5.17 & J0438+3004 & J0510+1800  \\
   8   & 398, 400, 410, 412 & 15 Jul. 2021 & 10.55 & J0438+3004 & J0423-0120 \\
\hline
  \end{tabular}
  \end{center}
\end{table*}

\subsection{Data reduction and imaging}
The observed visibilities were reduced and calibrated using the Common Astronomical Software Application (CASA) package \citep{McMullin+07}. 
The initial flagging of the visibilities and the calibrations for the bandpass characteristics, complex gain, and flux scaling were performed using the pipeline scripts provided by ALMA.
After flagging the bad data, the corrected data were concatenated and imaged by CLEAN. 
The CLEAN map was created by adopting Briggs weighting with a robust parameter of 0.5.
After that, three rounds of phase and one round of amplitude self-calibrations were performed.
The final images have a synthesized beam size $\Theta$ of 0$\farcs$085$\times$0$\farcs$045, 0$\farcs$095$\times$0$\farcs$054, 0$\farcs$099$\times$0$\farcs$062 and 0$\farcs$080$\times$0$\farcs$066, with a observing frequency of 138.0, 224.0, 336.5 and 398.0 GHz (correspponding to an observing wavelength of 2.17, 1.34, 0.89 and 0.75 mm) at Band 4, 6, 7 and 8, respectively.
The total flux density $F_{\nu}$, peak intensity $I_{\rm peak}$, RMS noise level and 67\% and 95\% emission radius are summarized in Table \ref{table:1}.

\begin{table*}[ht]
  \begin{center}
  \caption{Summary of observed data}
  \label{table:1}
  \begin{tabular}{ccccccccc}
  \hline
 Band & $\lambda$ & $\nu$ & $\Theta$ & $F_{\nu}$ & $I_{\rm peak}$ & noise & $r_{\rm d,67}$ & $r_{\rm d,95}$  \\
      & (mm) & (GHz) & (arcsec) & (mJy) & (${\rm mJy/beam}$) & (${\rm \mu Jy/beam}$) & (au) & (au)  \\
\hline\hline
 4 & 2.17  & 138.0 & 0$\farcs$085$\times$0$\farcs$045 & 21.3 & 2.52 & 6.67 & 30.0 & 46.0 \\
 6 & 1.34  & 224.0 & 0$\farcs$095$\times$0$\farcs$054 & 68.0 & 8.33 & 10.5 & 32.0 & 48.0 \\
 7 & 0.890 & 336.5 & 0$\farcs$099$\times$0$\farcs$062 & 154  & 18.7 & 66.7 & 34.0 & 52.0 \\
 8 & 0.753 & 398.0 & 0$\farcs$080$\times$0$\farcs$066 & 215  & 19.1 & 115 & 34.0 & 58.0 \\
\hline
  \end{tabular}
  \end{center}
\end{table*}

\section{Results} \label{sec:results}

\subsection{ALMA multi-band images of the CW Tau disk}
Figure \ref{fig:images} shows the obtained images of the CW Tau disk at ALMA Band 4, 6, 7 and 8.
The integrated flux density at Band 4, 6, 7 and 8 are 21.3, 68.0, 154 and 215 mJy, respectively.
The flux density at Band 7 is consistent with that previously reported by \citet{Bacciotti+18} within the error.
The 2D Gaussian fit of the intensity map at the four bands gives a disk position angle ${\rm PA}_{\rm disk}$ of 60.9--61.9$^{\circ}$ and a disk inclination $i_{\rm disk}$ of 57.8--59.2$^{\circ}$, almost independent on the images.
Both the obtained position angle and inclination are consistent with previous Band 7 observation (\citealt{Bacciotti+18}).
The obtained position angle is perpendicular to the geometrical configuration of a jet ejected from the central star (\citealt{Hartigan+04}).

In Band 4 and 6 images, we identify a gap-like structure at $\sim$ 20 au in the major-axis direction,  which is close to current Neptune orbit.
In the minor-axis direction the gap is not clearly detected probably because of the beam dilution.
The detailed structure of the gap will be discussed in Section \ref{sec:radial}.

\subsection{Spectral energy distribution}
\begin{figure}[ht]
\begin{center}
\includegraphics[scale=0.46]{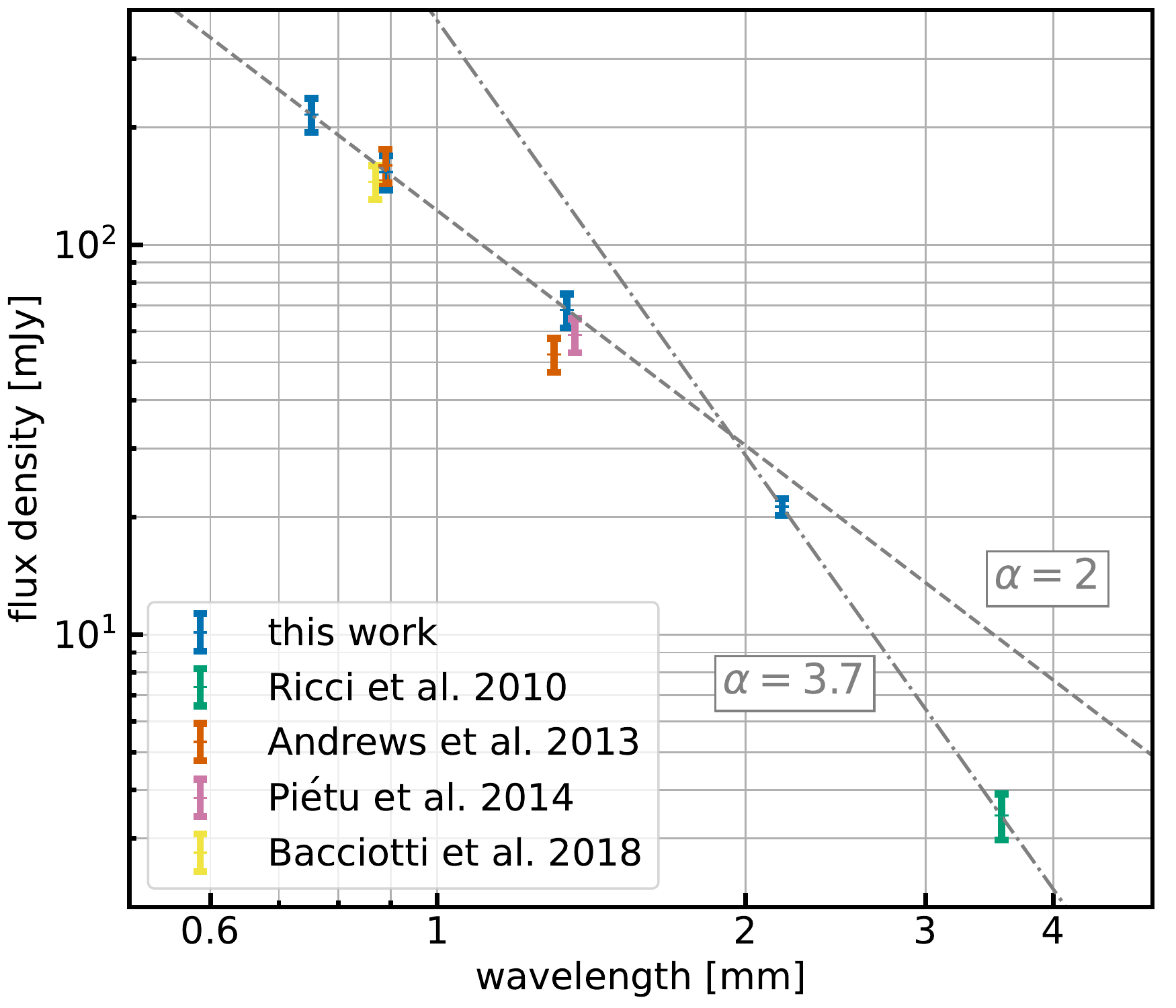}
\caption{
Millimeter SED of the CW Tau disk.
For comparison, we plot a spectral slope of 2 (anchored at the Band 8 flux density) and 3.7 (anchored at the Band 4 flux density) with a dashed and dash-dotted line, respectively.
}
\label{fig:sed}
\end{center}
\end{figure}

Figure \ref{fig:sed} shows the SED at millimeter wavelengths of the CW Tau disk.
In Figure \ref{fig:sed}, in addition to our data, we also plot data taken from literature \citep{Ricci+10,Andrews+13,Pietu+14,Bacciotti+18}.
For our data, we set a potential calibration error in the absolute flux as 5\% for ALMA Band 4 and 10\% for Band 6, 7 and 8, following the ALMA official observing guide.
We assume the error as 10\% of the absolute values for the flux densities taken from previous studies except for the 3.56 mm flux density.
For the 3.56 mm flux density, we set the error as 13.5\% which is a square root of sum of squares of the thermal noise and absolute flux uncertainty \citep{Ricci+10}.
Our flux density obtained at ALMA Band 6 is consistent, within the errors, with that at a similar wavelength obtained by \citet{Pietu+14} but slightly higher than that obtained by \citet{Andrews+13}.
The obtained Band 7 flux density is consistent with that of \citet{Andrews+13} as well as \citet{Bacciotti+18}.

The power-law fitting of our data yields a spectral index of $2.01\pm0.239$ between Bands 8 and 6, $2.400\pm0.231$ between Bands 6 and 4, $3.69\pm0.291$ between Band 4 and 3.56 mm.
The spectral index at shorter wavelengths is consistent with that of optically thick emission.
In contrast, the total flux density at 2.17 (ALMA Band 4) and 3.56 mm \citep{Ricci+10} are significantly lower than the flux density extrapolated from shorter wavelengths with a spectral slope of 2.
The spectral slope of $\sim$ 3.7 is consistent with that of optically thin emission from ISM-like grains smaller than $\lambda/2\pi\sim350~{\rm \mu m}$ (e.g., \citealt{FDS99,Draine06,Planck+14}).
Between these two regimes, Bands 6 and 4, the disk is expected to be partially optically thin and the spectral index is in the middle of the two regimes, $\alpha\sim2.4$.

\subsection{Radial intensity profiles} \label{sec:radial}
\begin{figure}[ht]
\begin{center}
\includegraphics[scale=0.46]{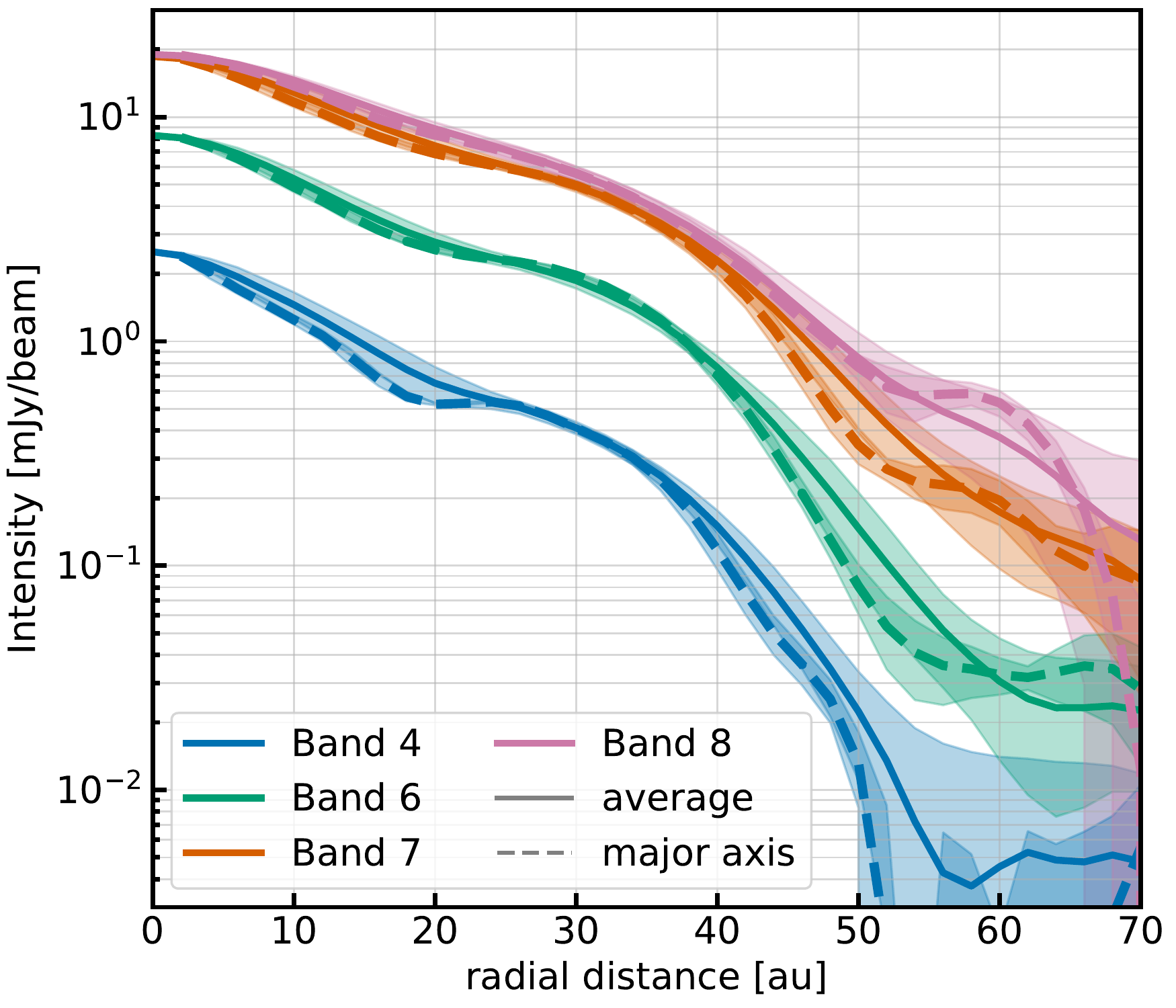}
\caption{
Radial intensity profile at ALMA Band 4, 6, 7 and 8.
The solid lines show the azimuthally averaged intensity and the dashed lines show the intensity averaged over the region $\pm7.5^{\circ}$ from the major axis.
The shaded region corresponds to the standard deviation of the intensity at each radial grid with an uniform radial width of 2 au.
}
\label{fig:radial}
\end{center}
\end{figure}
Figure \ref{fig:radial} shows the azimuthally averaged intensity profile at ALMA Band 4, 6, 7 and 8.
In the averaging, we assume $i_{\rm disk}=59^{\circ}$ and ${\rm PA}_{\rm disk}=61^{\circ}$, and the radial direction is uniformly separated into annuli with a width of 2 au.
We see that the intensity gradually decreases inside $\sim$ 40 au, while it steeply decreases outside $\sim$ 40 au.
To quantify the disk radius, we use the emission radius $r_{\rm d,67}$ and $r_{\rm d,95}$ within which 67\% and 95\% of the total emission originates.
We find that $r_{\rm d,67}$ is $\sim30$ au and almost independent on the observing wavelength; $r_{\rm d,67}=30.0$, 32, 34 and 34 au at Band 4, 6, 7 and 8, respectively.
If we extend the emission threshold up to 95\%, the emission radius is estimated as $r_{\rm d,98}=46$, 48, 52, 58 au at Band 4, 6, 7 and 8, respectively.
Although the emission radius could be used as a tracer of the dust-size segregation due to dust radial drift (e.g., \citealt{Rosotti+19}), the these emission radius does not clearly show the evidence of the dust-size segregation.

In contrast to the intensity averaged over the whole azimuthal direction, the intensity profile shows a gap at $\sim$ 20 au in Band 4 and 6 images when it is averaged over only major axis direction ($\pm7.5^{\circ}$ from the major axis).
Here we evaluate the gap structure in the Band 4 image quantitatively.
As seen in Figure \ref{fig:radial}, the gap structure is smeared out in the minor axis direction because the observing beam is elongated to that direction.
Therefore, we use the radial intensity profile around the major axis to estimate the gap structure.
To quantify the gap structure, we follow the approach used in \citet{Zhang+18}.
In this approach we evaluate the radial gap position $r_{\rm gap}$, radial ring position just outside the gap $r_{\rm ring}$, depth of the gap $\delta_{I}$ and width of the gap $\Delta_{I}$.
The gap and ring position is defined as the location where the radial intensity profile becomes local minimum and maximum, respectively.
We obtain $r_{\rm gap}=20$ au and $r_{\rm ring}=24$ au.
The gap depth is defined as $\delta_{I}\equiv I_{\nu}(r_{\rm ring})/I_{\nu}(r_{\rm gap})$ and is evaluated as 1.025.
The gap width is defined as $\Delta_{I}\equiv(r_{\rm out}-r_{\rm in})/r_{\rm out}$, where $r_{\rm in}$ and $r_{\rm out}$ are the radial location where the radial intensity profile reaches $I_{\nu}=\{I_{\nu}(r_{\rm ring})+I_{\nu}(r_{\rm gap})\}/2$ with $r_{\rm in}$, $r_{\rm out}<r_{\rm ring}$.
We derive $r_{\rm in}$ and $r_{\rm out}$ as 18 and 22 au, which results in $\Delta_{I}=0.18$.
The gap has a radial width of 4 au, much smaller than the spatial resolution of our observation ($\sim$ 8.2 au for $d=132$ pc), indicating that the gap is not spatially resolved with the current resolution.

\section{Disk modelling} \label{sec:diskmodelling}
In this section, to take a more in-depth look at dust properties, we characterize the CW Tau disk using the spatially-resolved ALMA observations at Band 4, 6, 7 and 8.

\subsection{Methodology}
To characterize the dust disk structure, we employ multi-band analysis used in \citet{Macias+21} and \citet{Sierra+21}.
In this approach, we model the intensity at each radius with a parameter set of the dust temperature $T$, dust surface density $\Sigma_{\rm d}$ and maximum dust radius $a_{\rm max}$.
Given these three paramters, the emergent specific intensity is calculated as \citep{Sierra+19,Carrasco+19}
\begin{eqnarray}
I_{\rm \nu}=B_{\rm \nu}(T)\left\{ 1-\exp{\left(-\frac{\tau_{\nu}}{\mu}\right)} +\omega_{\nu} F(\tau_{\nu},\omega_{\nu})\right\},
\label{eq:intensity}
\end{eqnarray}
where $B_{\nu}(T)$ is the Planck function with temperature $T$, $\tau_{\nu}$ is the vertical extinction optical depth, $\omega_{\nu}$ is the effective scattering albedo and $\mu\equiv\cos{i}$ represents the effect of the disk inclination.
The effect of scattering emerges from $F(\tau_{\nu},\omega_{\nu})$ given as
\begin{eqnarray}
F(\tau_{\nu},\omega_{\nu})=\frac{f_{1}(\tau_{\nu},\omega_{\nu})+f_{2}(\tau_{\nu},\omega_{\nu})}{\exp{(-\sqrt{3}\epsilon_{\nu}\tau_{\nu})}(\epsilon_{\nu}-1)-(\epsilon_{\nu}+1)}
\label{eq:Ffactor}
\end{eqnarray}
where
\begin{eqnarray}
f_{1}(\tau_{\nu},\omega_{\nu})=\frac{ 1-\exp{\left\{ -(\sqrt{3}\epsilon_{\nu}+1/\mu)\tau_{\nu}\right\}} }{\sqrt{3}\epsilon_{\nu}\mu+1}
\end{eqnarray}
and
\begin{eqnarray}
f_{2}(\tau_{\nu},\omega_{\nu})=\frac{\exp{\left(-\tau_{\nu}/\mu\right)}-\exp{(-\sqrt{3}\epsilon_{\nu}\tau_{\nu})}}{\sqrt{3}\epsilon_{\nu}\mu-1},
\end{eqnarray}
with $\epsilon_{\nu}\equiv\sqrt{1-\omega_{\nu}}$.

\begin{figure}[ht]
\begin{center}
\includegraphics[scale=0.55]{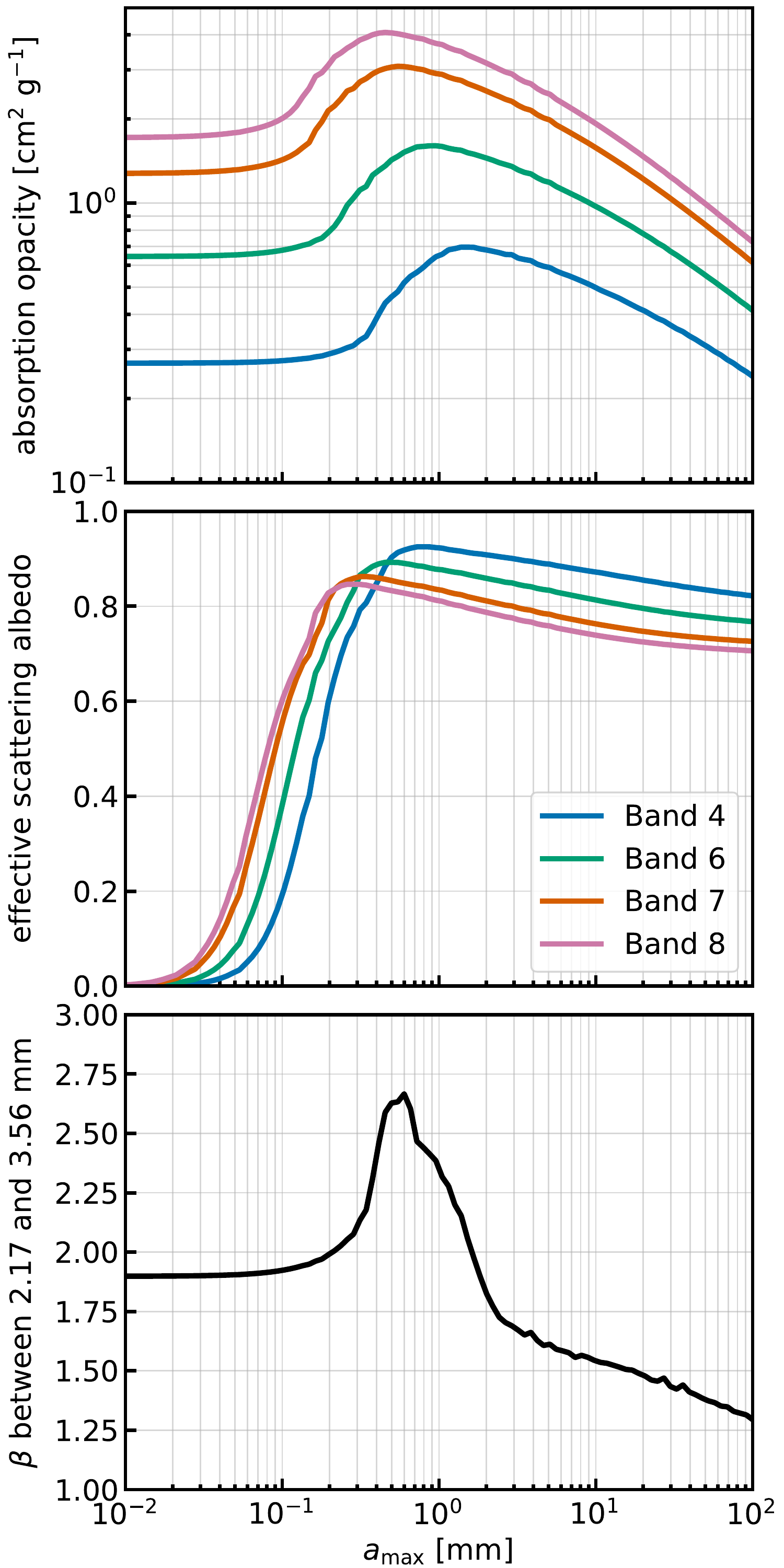}
\caption{
Apsorption opacity (top) and effective scattering albedo (middle) of our dust model at ALMA Band 4, 6, 7 and 8.
The bottom panel shows the opacity slope between 2.17 and 3.56 mm.
}
\label{fig:opac}
\end{center}
\end{figure}

For the dust opacities, we consider a compact spherical dust with a size distribution ranging from $0.05~{\rm \mu m}$ to $a_{\rm max}$ with a slope of $-3.5$.
The dust composition is assumed to be that of DSHARP model (\citealt{Birnstiel+18}; see also \citealt{HS96,Draine06,WB08} for each dust component) and opacities are calculated with Optool \citep{optool} with a DHS factor of 0.8 which mimics the irregularity of the dust shape \citep{MdK05}.
The dust-size distribution is divided into 300 bins to compute the size-averaged opacities.
Given the dust model, the absorption and scattering opacities are determined only by $a_{\rm max}$.
The obtained absorption opacity, effective scattering albedo and opacity slope $\beta$ are shown in Figure \ref{fig:opac}.

Once we obtain the model intensities at each radius and at each wavelength, we adopt a Bayesian approach to obtain the posterior probability distributions of the model parameters with using a standard likelihood function:
\begin{eqnarray}
P(I_{\rm B4},I_{\rm B6},I_{\rm B7},I_{\rm B8}|T,\Sigma_{\rm d},a_{\rm max}) \propto \exp{\left(-\frac{\chi^{2}}{2}\right)},
\end{eqnarray}
where
\begin{eqnarray}
\chi^{2}=\sum_{i}\left(\frac{I_{{\rm obs},i}-I_{{\rm m},i}}{\sigma_{i}}\right)^{2},
\end{eqnarray}
where $I_{{\rm obs},i}$ and $I_{{\rm m},i}$ are observed and model intensity at frequency $i$, respectively.
The uncertainty in the observed intensity at frequency $i$, $\sigma_{i}$, is given as
\begin{eqnarray}
\sigma_{\rm i}^{2} = \Delta I_{{\rm obs},i}^{2} + (\delta_{\rm i} I_{{\rm obs},i})^{2},
\end{eqnarray}
where $\Delta I_{{\rm obs},i}$ is the standard deviation of the azimuthally averaged intensity at frequency $i$ and $\delta_{i}$ represents the absolute flux calibration error at frequency $i$.
We set $\delta_{i}$ as 5\% for ALMA Band 4 and 10\% for Band 6, 7 and 8, following the ALMA official observing guide.

We vary $T_{\rm d}$, $\Sigma_{\rm d}$, and $a_{\rm max}$ in a $100\times200\times100$ grid, respectively. 
$T_{\rm d}$ is uniformly separated from 4 to 60 K.
$\Sigma_{\rm d}$ and $a_{\rm max}$ are varied in logarithmic space from $10^{-3}$ to $10^{3}$ ${\rm g~cm^{-2}}$, and from $10^{-3}$ to 10 cm, respectively. 
As previous studies have shown, this fitting approach usually yields two families of solutions: small grains (low scattering albedo) with low temperature and large grains (high scattering albedo) with high temperature.
Therefore, we fit the observations with two different dust-size regimes: from 10 to 300 ${\rm \mu m}$ (small grain model) and from 300 ${\rm \mu m}$ to 10 cm (large grain model).

Since the original images shown in Figure \ref{fig:images} have different angular resolutions, we produce new images with a common resolution of $0\farcs1\times0\farcs1$ with the CASA task imsmooth and then obtain azimuthally averaged intensity profiles which are used for the fitting.

\begin{figure*}[ht]
\begin{center}
\includegraphics[scale=0.5]{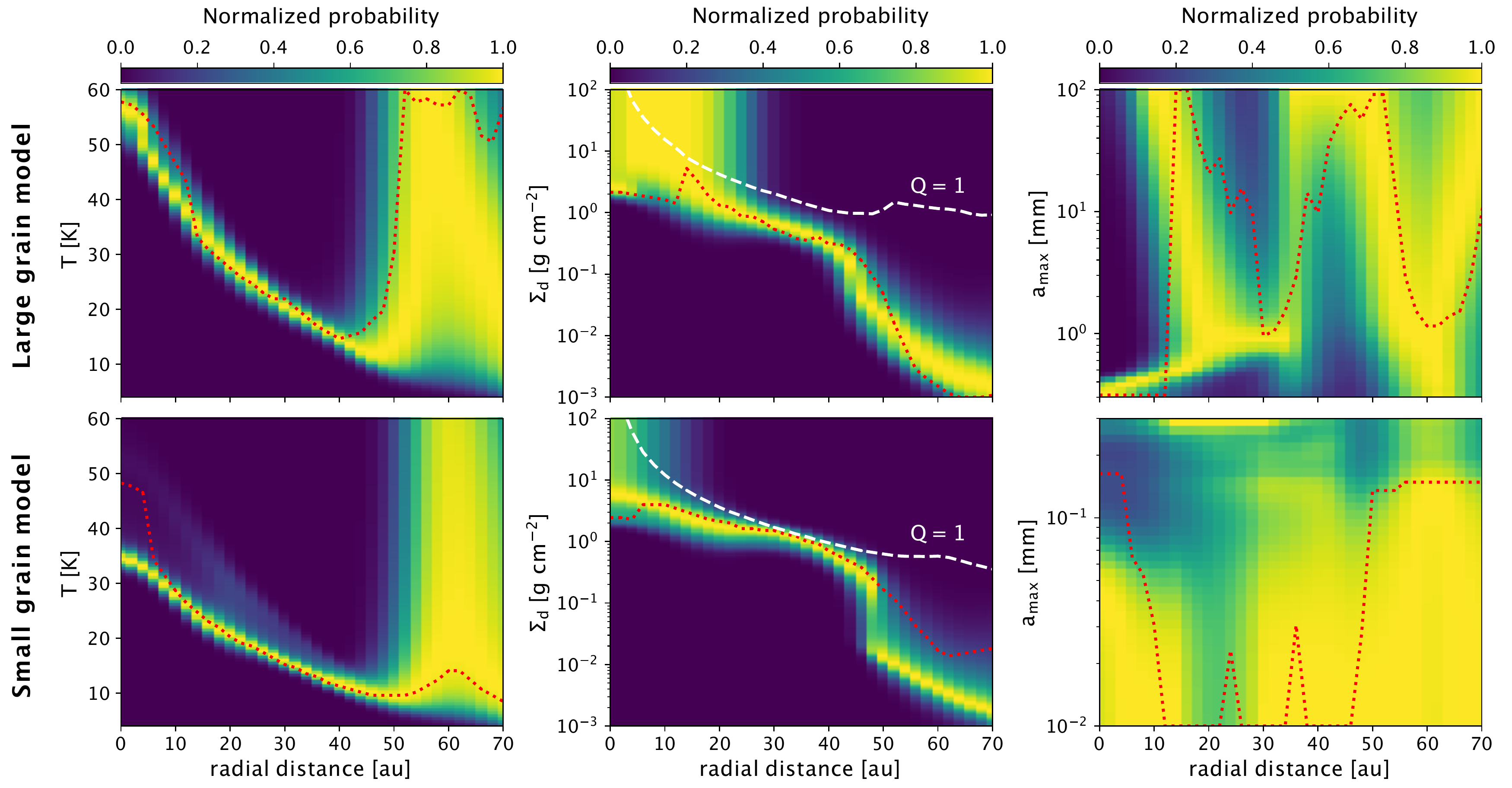}
\caption{
{\it top}: Marginalized posterior probability distributions of the model parameters fitted to the multi-band observations. The maximum dust size $a_{\rm max}$ varies from 300 ${\rm \mu m}$ to 10 cm.
The marginalized posterior probability is normalized by the maximum at each radius.
The white dashed line in the middle panel shows the dust surface density profile for which the disk is gravitationally unstable (i.e., $Q=1$), assuming a gas to dust ratio of 100.
The red dotted line shows the value that maximize the probability.
{\it bottom}: Same as top panels, but $a_{\rm max}$ varies from 10 to 300 ${\rm \mu m}$.
}
\label{fig:model}
\end{center}
\end{figure*}

\subsection{Modelling results} \label{sec:modelresult}
\subsubsection{Fitting with ALMA observations} \label{sec:fitALMA}
\begin{figure}[ht]
\begin{center}
\includegraphics[scale=0.46]{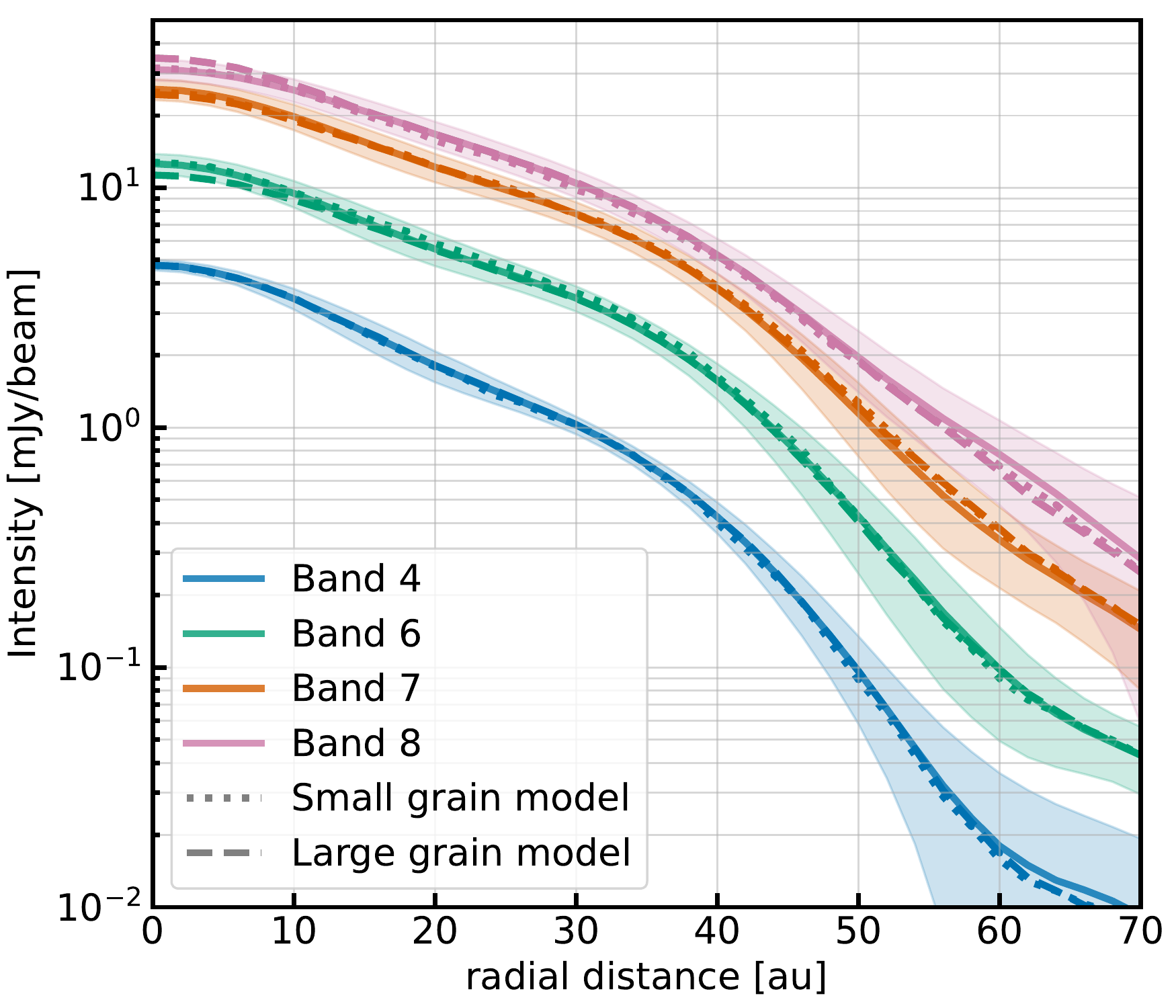}
\caption{
Comparison between the observed intensities and model intensities.
The solid lines denote the observed intensity with uncertainty due to thermal noise as well as the flux calibration error shown as shaded region.
The dotted and dashed lines show the model intensity derived from the small and large grain models, respectively.
}
\label{fig:comparison}
\end{center}
\end{figure}

Figure \ref{fig:model} shows the marginalized probability of each fitting parameters for large and small grain models.
The marginalized probability of a given parameter is obtained by integrating the probability over the other two parameters, and then normalized by the maximum value.
Figure \ref{fig:model} also shows the best model (shown in red solid line) which maximize the probability (not the marginalized probability).
The intensity profiles obtained by the best models are compared with the observations in Figure \ref{fig:comparison}.
We clearly see that both large and small grain models well reproduce the observed intensity within the errors.

In each model, the temperature profile within $\sim$ 50 au is well constrained.
The large grain model predicts higher temperature  than the small grain model because large grains have large scattering albedo which reduces the emergent intensity.
At outer region ($\gtrsim 50$ au), the dust temperature is not well constrained because the signal-to-noise ratio is too low due to steep decrease in the observed intensity beyond 50 au.
Although the temperature is not well constrained, the dust surface density is relatively well constrained and shows a tail-like structure with a transition at $\sim$ 50 au.
The presence of dust grains at outer region can be inferred from the radial intensity profile at Band 6, 7 and 8 (see Figure \ref{fig:comparison}).
However, the signal-to-noise ratio is $\sim$ 2--3  at 50--70 au at those wavelengths, which is not enough to characterize the outer disk properties in detail.

\begin{figure}[ht]
\begin{center}
\includegraphics[scale=0.46]{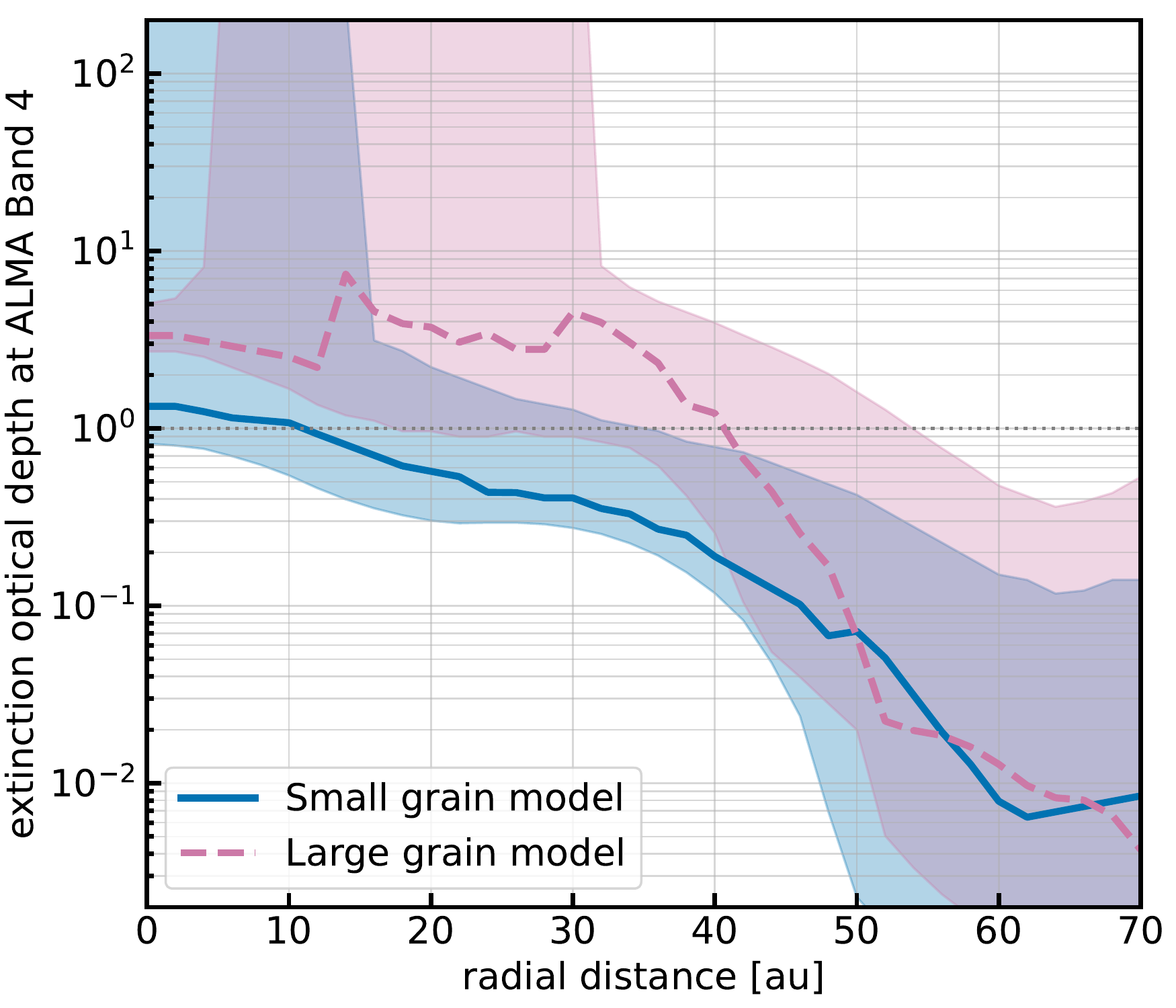}
\caption{
Extinction optical depth at ALMA Band 4 estimated from the fitting.
The blue solid and pink dashed lines denote the best solutions of small and large grain models.
The shaded region shows the 1-$\sigma$ confidence interval of the models.
}
\label{fig:tau}
\end{center}
\end{figure}

Figure \ref{fig:tau} shows the extinction optical depth at ALMA Band 4 of the best models with 1-$\sigma$ confidence interval in the models.
The 1-$\sigma$ confidence interval is evaluated with the criterion:
$\Delta \chi^{2}\equiv \chi^{2}-\chi_{\rm min}^{2}<3.53$ where $\chi_{\rm min}^{2}$ is $\chi^{2}$ of the best solution (i.e., minimum value of $\chi^{2}$). The critical value of 3.53 comes from the fact that we use three parameters for the fitting (e.g., \citealt{Press+07}).
And then, we search the maximum and minimum optical depth in all models with $\Delta \chi^{2}<3.53$, and plot it in Figure \ref{fig:tau} as the 1-$\sigma$ confidence interval.
In the large grain model, the disk is fully optically thick inside $\sim$ 30 au at ALMA Band 4, and hence only lower limit of the surface density is obtained.
A similar behavior is also seen in the small grain model inside $\sim$ 15 au, but the dust surface density is relatively well constrained compared to the large grain model because small grains have small scattering albedo (see Figure \ref{fig:opac}).

\begin{figure}[ht]
\begin{center}
\includegraphics[scale=0.46]{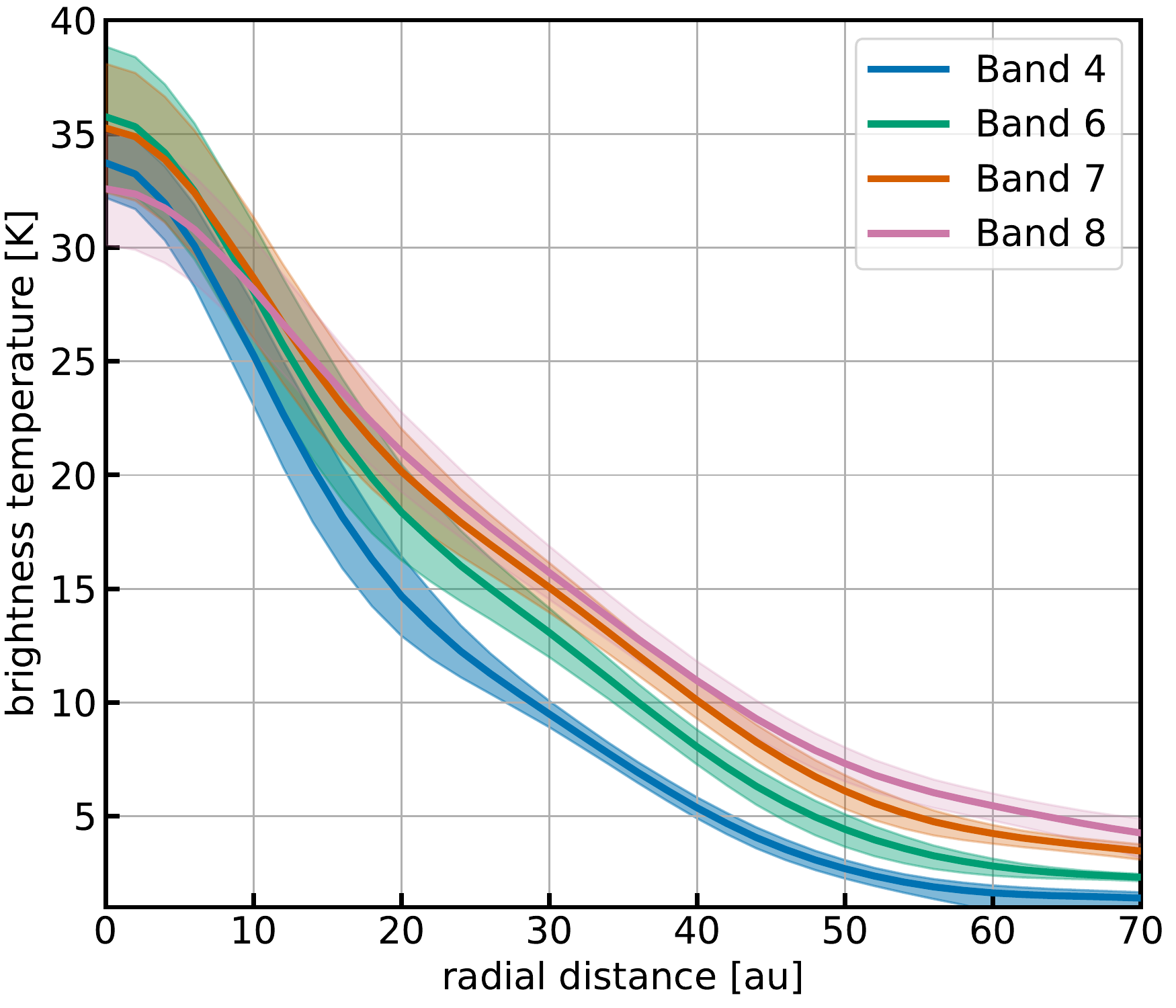}
\caption{
Brightness temperature profile of the CW Tau disk at ALMA Band 4, 6, 7 and 8.
The brightness temperature is derived from the images with a common beam size of 0.1 arcsec (13 au in diameter).
The shaded region corresponds to the potential error caused by the flux calibration as well as the thermal noise.
}
\label{fig:Tb}
\end{center}
\end{figure}

In the large grain model, the maximum dust size is preferred to be $\sim300~{\rm \mu m}$ inside 10 au, while it is larger beyond 10 au.
This is because the observed intensity at ALMA Band 4 is high compared to the other bands.
To reproduce the high intensity at Band 4, the scattering albedo needs to be low at Band 4 and hence smaller grains are preferred.
A similar behavior is also seen in the temperature profile; the best temperature model steeply increases inside 10 au to account for the high intensity at Band 4.
This might be because of the mid-plane heating which is not considered in our fitting approach.
If the disk is optically thick, the longer observing wavelength traces closer to the mid-plane which has higher temperature due to the mid-plane heating.
To see this point, we show the observed brightness temperature in Figure \ref{fig:Tb}.
At $\lesssim10$ au, the brightness temperatures at Band 6 and 7 are higher than that at Band 8, which is unlikely if the disk is vertically isothermal.
However, the difference is within 1-$\sigma$ uncertainty and hence it is unclear if the midplane heating indeed affect the observed intensities.
It would be worth noting that if the inner region is optically thick, we can only observe an upper layer and hence larger grain may exist at the midplane (\citealt{Sierra+20,Ueda+21b}).

At $\sim20$ au, the maximum dust size is preferred to be large ($\gtrsim 500~{\rm \mu m}$) and not well constrained.
This is because of unresolved gap structure found in original images (Figure \ref{fig:images}).
A gap at $\sim$ 20 au in Band 4 image is smeared out in the radial intensity profile used for the fitting.
This unresolved gap lowers the radial intensity at $\sim 20$ au at Band 4.
To explain the low intensity at Band 4, dust grains need to have high scattering albedo at Band 4 and hence large grains $(\gtrsim 300~{\rm \mu m})$ are preferable (see Figure \ref{fig:opac}).
This behavior is also seen in the small grain model where the largest grain is desired.
This means that the preferred dust size around 20 au is significantly affected by the beam dilution and hence the true dust size may differ from the model predictions.
In the small grain model, the maximum dust size is not well constrained except at $\sim20$ au.
This is because, in the regime of $a_{\rm max}\lesssim100~{\rm \mu m}$, the intensity is not sensitive to the dust size because absorption opacity is independent on the size and scattering is not effective.

In the middle panels of Figure \ref{fig:model}, we plot the dust surface density above which the disk is gravitationally unstable, derived from the criterion \citep{Toomre64}:
\begin{eqnarray}
Q\equiv\frac{c_{\rm s}\Omega_{\rm K}}{\pi G \Sigma_{\rm g}} =1,
\label{eq:Q}
\end{eqnarray}
where $c_{\rm s}$ is the sound speed of the gas, $\Omega_{\rm K}$ is the Keplerian rotational frequency, $G$ is the gravitational constant and $\Sigma_{\rm g}$ is the gas surface density.
We assume a constant dust-to-gas mass ratio of 0.01 for simplicity.
The stellar mass of CW Tau is set as $1.6M_{\odot}$ \citep{Andrews+13}.
For the sound speed, we use the best model of the temperature profile, denoted by the red dotted line in the right panels of Figure \ref{fig:model}.
Interestingly, our best dust surface density curve reaches $Q\sim$ 1--3 at most of the outer region ($\sim$ 20--50 au).
This suggests that CW Tau might be marginally gravitationally unstable, although the dust-to-gas mass ratio has a large uncertainty.

\subsubsection{Comparison with the 3.56 mm flux density} 
\begin{figure}[ht]
\begin{center}
\includegraphics[scale=0.46]{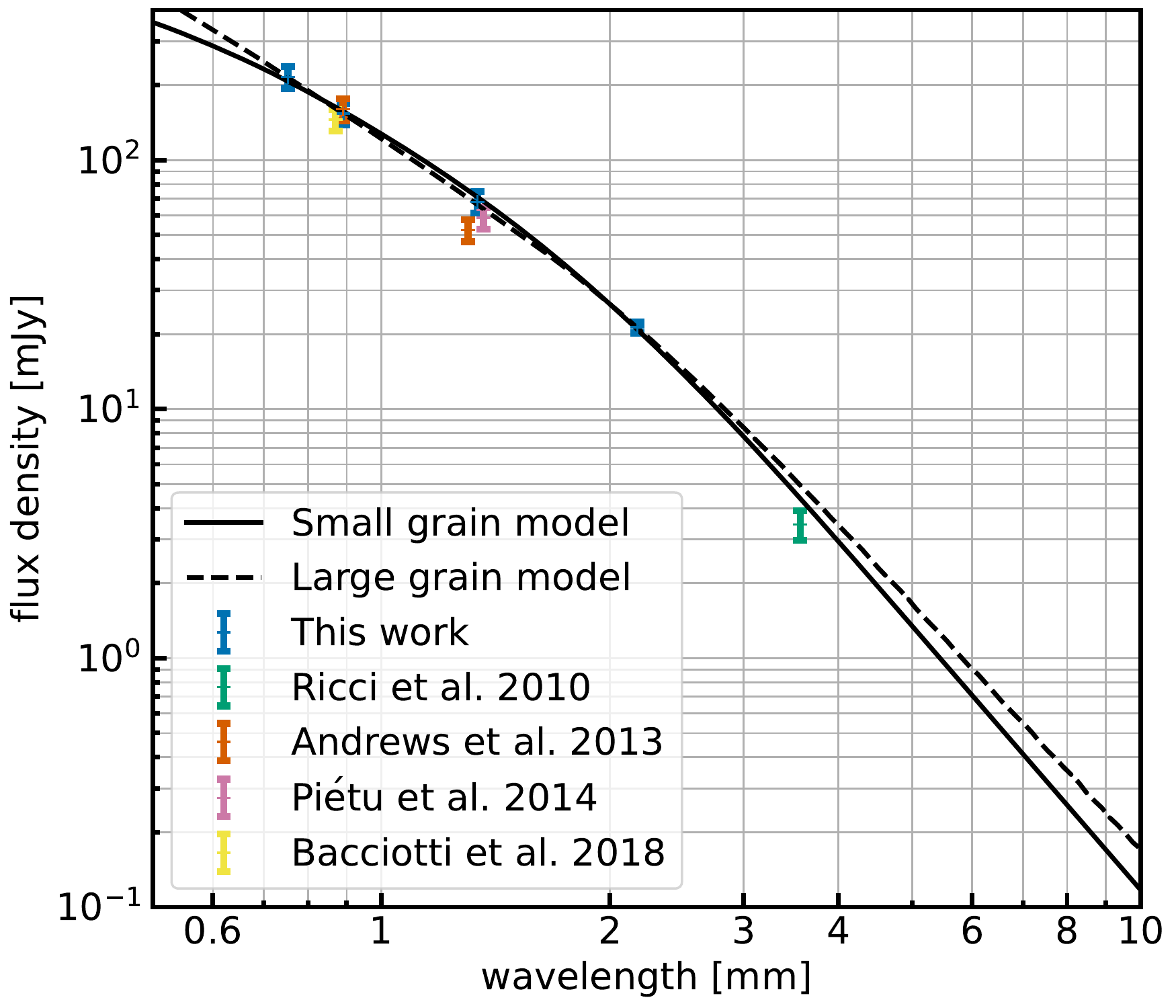}
\caption{
Comparison of the observed spectral energy distribution with the best solution in the small and large grain models
}
\label{fig:SED_predict}
\end{center}
\end{figure}

In Section \ref{sec:fitALMA}, the parametric fitting is performed for ALMA Band 4, 6, 7 and 8 data.
However, as shown in Figure \ref{fig:sed}, the disk is expected to be optically thick except at Band 4 and hence it is difficult to obtain the spectral slope at optically thin regime (i.e., opacity slope) from these observations.
In this point, the flux density at 3.56 mm is crucial to constrain the dust size in the CW Tau disk.
Figure \ref{fig:SED_predict} compares the observed SED with our best models of small and large grains. 
We see that the model SEDs well reproduce the observed flux density at ALMA bands.
However, both models predict higher flux density at 3.56 mm than the observation.
To constrain the dust size in more detail, we investigate the impact of the dust size on the SED particularly at 3.56 mm.
Since the 3.56 mm observation does not spatially resolve the disk, we perform a parametric fitting for the ALMA observations with $a_{\rm max}$ being uniform in the entire region.

\begin{figure}[ht]
\begin{center}
\includegraphics[scale=0.42]{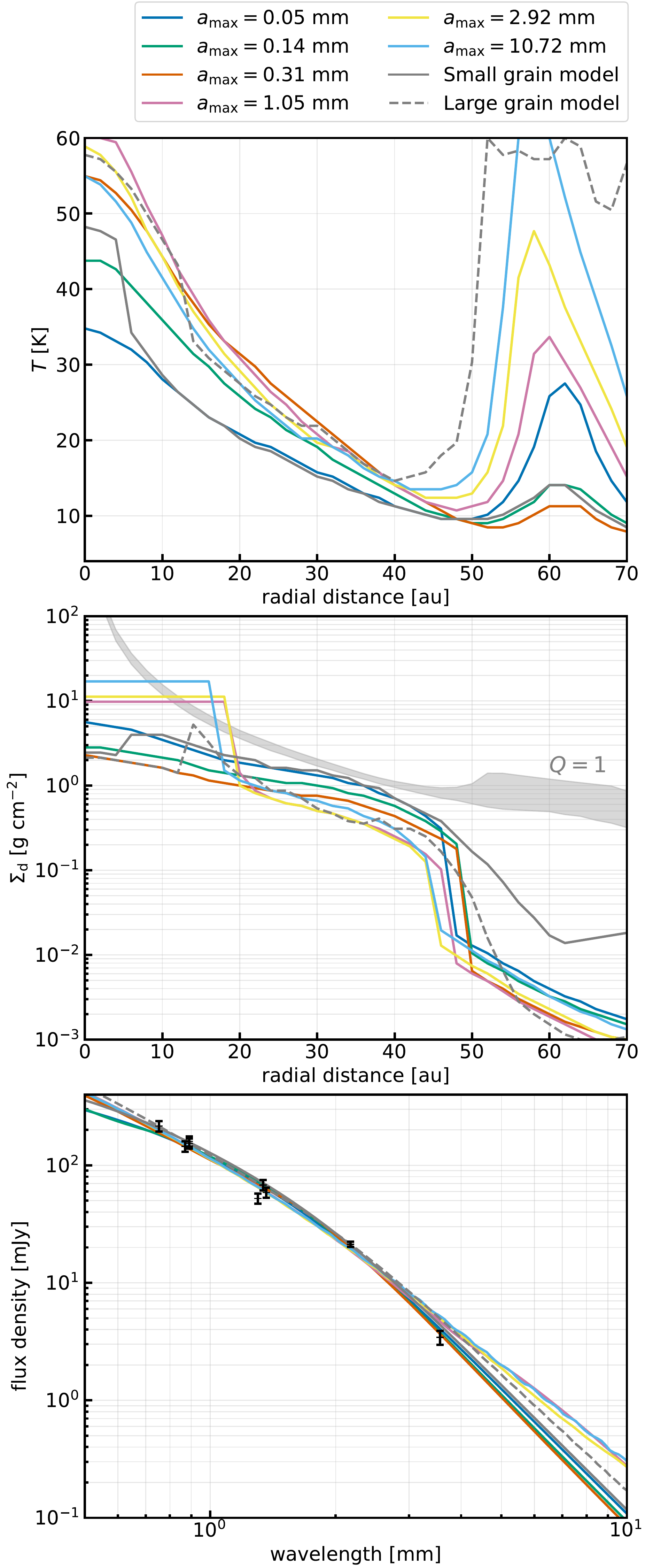}
\caption{
The best temperature (top) and surface density (middle) profiles for models with a fixed dust size.
The gray-colored region in the middle panel denotes the condition for $Q=1$ assuming a dust-to-gas mass ratio of 0.01 and temperature profiles shown in the top panel.
The obtained SEDs are shown in the bottom panel.
We also plot the small and large grain models shown in Figure \ref{fig:model} for comparison.
}
\label{fig:diskmodel_com}
\end{center}
\end{figure}

Figure \ref{fig:diskmodel_com} shows the best temperature and dust surface density profiles and resultant SED for each maximum dust size.
Generally, in $a_{\rm max}\lesssim1$ mm, the predicted temperature increases with $a_{\rm max}$ since larger grains has larger scattering albedo and hence need higher temperature to obtain the given (observed) intensity at optically thick regime.
In contrast,  in $a_{\rm max}\gtrsim 1$ mm, the predicted temperature decreases with increasing $a_{\rm max}$ since scattering albedo decreases with increasing the dust size (see Figure \ref{fig:opac}).
The predicted surface density reflects the dependence of the temperature on the dust size: dust surface density needs to be higher for lower temperature to obtain the given observed intensity.
For the models with $a_{\rm max}\gtrsim1$ mm, the dust surface density is not constrained in $r\lesssim20$ au, since the disk is optically thick even at ALMA Band 4 as mentioned in Section \ref{sec:fitALMA}.
The predicted surface density has a steep cutoff at $\sim 50$ au independent on the dust size.

\begin{figure}[ht]
\begin{center}
\includegraphics[scale=0.46]{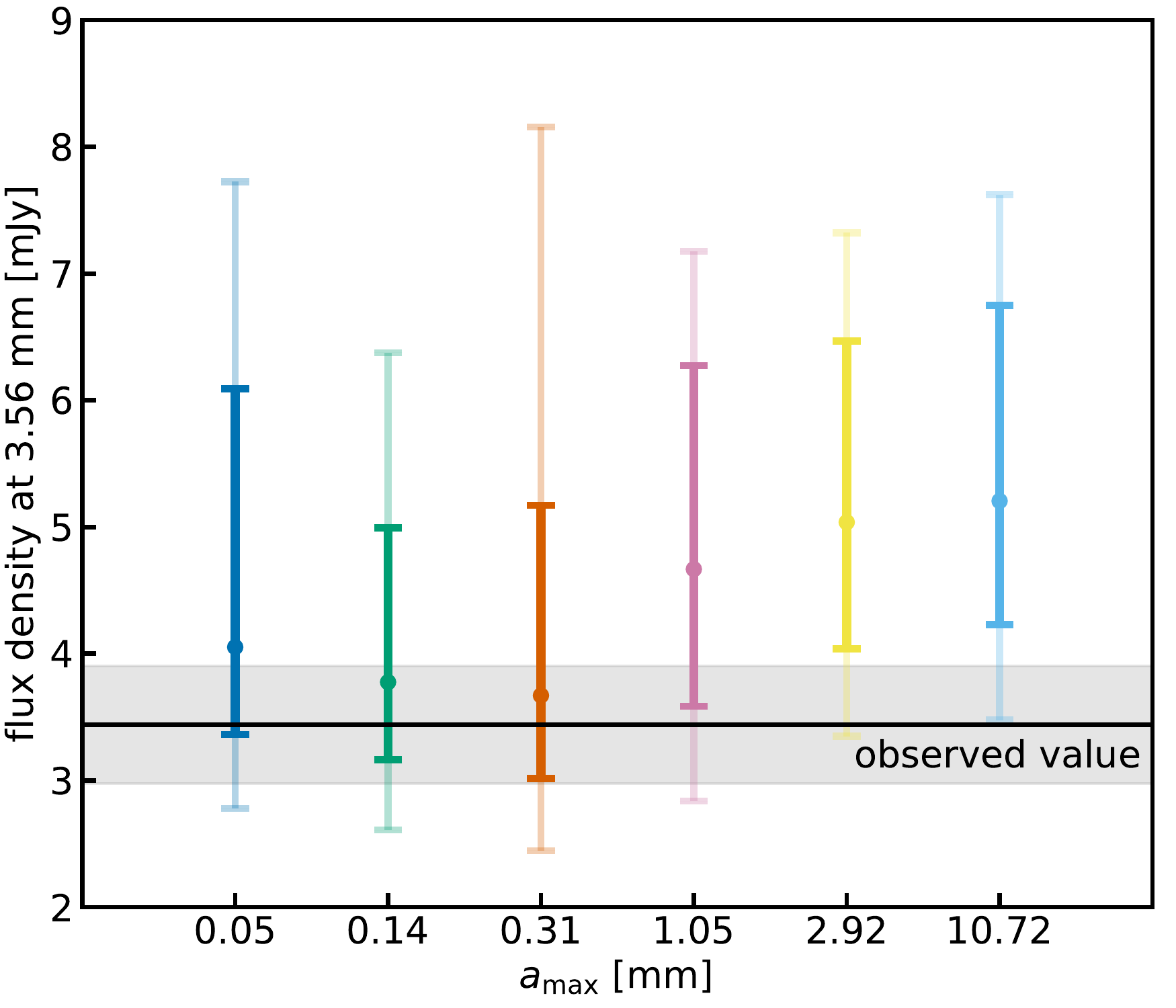}
\caption{
Comparison of the observed 3.56 mm flux density \citep{Ricci+10} and model predictions.
The thick and thin errorbars denote the 1- and 2-$\sigma$ confidence intervals in the models.
}
\label{fig:SED_error}
\end{center}
\end{figure}

We can clearly see that all models well reproduce the observed SED at ALMA wavelengths that are used for the fitting.
However, at 3.56 mm, models with $a_{\rm max}\gtrsim1$ mm predicts higher total flux density than the observation.
To take a close look at the 3.56 mm flux density, we show the obtained 3.56 mm flux density and 1- and 2-$\sigma$ confidence intervals in the models in Figure \ref{fig:SED_error}.
To obtain the confidence intervals of the models, for each radial grid, we calculate the maximum (minimum) intensity that reproduces the observations with 1- and 2-$\sigma$ confidence which are evaluated with $\Delta \chi^{2}<2.3$ and 6.2, respectively (e.g., \citealt{Press+07}). 
The maximum (minimum) flux density is calculated with integrating the maximum (minimum) intensity over the whole disk.

We see that the maximum dust size of $\lesssim1$ mm can reproduce the observed 3.56 mm flux with 1-$\sigma$ confidence, while large grains ($\gtrsim1$ mm) overestimates.
This is because the spectral slope between 2.17 and 3.56 mm is very steep ($\alpha\sim3.7$).
To explain the high spectral index, assuming the Rayleigh--Jeans limit, the opacity slope $\beta$ needs to be larger than $3.7-2=1.7$, which is satisfied only when $a_{\rm max}\lesssim 2$ mm (see Figure \ref{fig:opac}).
Furthermore, a disk with mm-sized grain needs to be optically thick at 2.17 mm, and hence spectral index cannot be as large as 3.7.
It should be noted that, although the steep spectral index between 2.2 and 3.6 mm prefers small grains ($\lesssim 1$ mm) with 1-$\sigma$ uncertainty, large grains ($\gtrsim 1$ mm) are still acceptable with 2-$\sigma$ uncertainty. 
To solve the degeneracy, it is important to use either high resolution ALMA Band 3 observation which would improve the model accuracy or observations at longer wavelengths, e.g., VLA, where the flux is more sensitive to the dust size.

\section{Discussion} \label{sec:discussion}

\subsection{Dust properties} \label{sec:dustproperties}
Previous ALMA polarimetric observation at Band 7 has shown the scattering-induced polarization pattern with a polarization degree of $\gtrsim$ 1\% \citep{Bacciotti+18}.
Since the scattering-induced polarization is effective only when $\lambda\sim a_{\rm max}/2\pi$ \citep{Kataoka+15}, the maximum grain size is expected to be $\sim$ 140 ${\rm \mu m}$ (see \citealt{Bacciotti+18}).
Furthermore, the high polarization degree of the CW Tau disk favors compact dust grains because porous dust aggregates cannot polarize the thermal emission \citep{Tazaki+19}.
The large spectral index between 2.17 and 3.56 mm indicates that the maximum dust size is smaller than mm, which is consistent with the dust size inferred from the previous polarimetric observation.
It would be worth noting that, however, non-spherical grains can produce higher polarization degree than spherical ones in the regime of $\lambda\lesssim a_{\rm max}/2\pi$ \citep{KB20}. Therefore, we cannot fully rule out the possibility that mm-sized or larger grains exist in the CW Tau disk.
Furthermore,scattering-induced polarization degree can be high even if the maximum dust size is larger than the observing wavelength when large grains settle to the mid-plane where ALMA Band 7 cannot observe \citep{Ueda+21b}.

If the dust size is indeed small, the dust disk is quite massive and the gas disk might be marginally gravitationally unstable, although it shows no clear signature of gravitational instability.
Similar behavior has been also found in the other disks \citep{Macias+21,Sierra+21}.
This anomalously high dust mass implies that the dust size is not small or we miss something in our model.
One potential solution is a carbonaceous material in the dust composition, which has a high absorption efficiency (e.g., \citealt{Zubko+96}).
The frequently used DSHARP opacity does not include the carbonaceous material.
The dust absorption opacity can be higher, if organic materials in the DSHARP dust is exchanged for the carbonaceous material, which lowers the estimated disk mass \citep{Birnstiel+18}.

\subsection{Disk mass estimates} \label{sec:diskmass}
In this subsection, we estimate the disk mass with different approaches.
First, let us estimate the disk mass using the observed total flux density.
Assuming dust emission at mm wavelengths is isothermal and optically thin, the total flux density at a given wavelength can be converted into the mass of the emitting dust $M_{\rm dust}$ with an equation \citep{Hildebrand+83}:
\begin{eqnarray}
M_{\rm dust} = \frac{F_{\nu}d^{2}}{\kappa_{\nu}B_{\nu}(T_{\rm dust})}.
\label{eq:diskmass}
\end{eqnarray}
Although the dust opacity has a large uncertainty, we consider a compact spherical dust with radius of $140~{\rm \mu m}$, which is inferred from the previous polarimetric observation as discussed in Section \ref{sec:dustproperties}.
Assuming $T_{\rm dust}=20$ K, which has been used in the literature (e.g., \citealt{Ansdell+16,Ansdell+18,Tazzari+21}), we obtain $M_{\rm dust}$ as 216, 115, 60 and 43 for the flux density at Band 4, 6, 7 and 8, respectively.
We obtain the disk mass as $M_{\rm dust}=229M_{\oplus}$ using the total flux density at 3.56 mm obtained by \citet{Ricci+10}.
The disk mass is lower at shorter observing wavelength because the assumption of optically thin limit is broken at shorter wavelengths.
Since the disk is expected to be optically thin at wavelengths of 2.17 mm (Band 4) and longer, the disk mass is converged into $\sim 230M_{\oplus}$ at those wavelengths.

The clear dependence of inferred dust mass on the observing wavelength gives us an important insight on the mass budget problem in planet formation where observed mean disk mass is significantly lower than the mean mass of observed planetary systems (e.g., \citealt{Ansdell+16,Manara+18,Cieza+19}).
Although the direct conversion using Equation \eqref{eq:diskmass} assumes optically thin regime, many of the previous estimates on disk masses rely on ALMA Band 6 or 7 where disks are potentially optically thick.
Our dust mass inferred from the flux density at ALMA Band 7 (0.89 mm) is $\sim3.8$ times smaller than that evaluated at 3.56 mm.
For reference, if we adopt a opacity of $\kappa=10(\nu/1000{\rm GHz})~{\rm cm^{2}~g^{-1}}$ that has been frequently used in the disk-mass surveys (e.g., \citealt{Ansdell+16,Tazzari+21}), we obtain $36M_{\oplus}$ at Band 6 which is significantly lower than that obtained from the multi-wavelengths analysis.

Furthermore, our analysis shows that ALMA Band 4 data is crucial for the dust-mass estimate.
Our SED analysis shows that the spectral index between ALMA Band 6 and 3.56 mm is $\sim 2.7$, while it is $\sim 3.7$ between Band 4 and 3.56 mm.
The low spectral index between Band 6 and 3.56 mm infers an optically thin emission of large grains (i.e., large absorption opacity), while the new Band 4 data reveals the disk is optically thick at Band 6 and would be composed of small grains (i.e., small absorption opacity).
Therefore, future survey study of the spectral index between ALMA Band 3 and 4 would be crucial for understanding the true disk mass.

The disk mass is also evaluated from our parametric fitting.
The small grain model (varing $a_{\rm max}$ with $r$) yields the best disk mass of 378$M_{\oplus}$, and 138$M_{\oplus}$ as a minimum mass in the 1-$\sigma$ confidence interval.
For the large grain model, the best disk mass is obtained as 230$M_{\oplus}$ and the minimum mass in the 1-$\sigma$ confidence interval is 80$M_{\oplus}$.
The maximum mass is not constrained as the disk is fully optically thick at Band 4 (see Figure \ref{fig:tau}).
When $a_{\rm max}$ is fixed for the entire region of the disk, the total dust mass is estimated as 344, 249 and 188$M_{\oplus}$ for models with $a_{\rm max}$ of  0.05, 0.14 and 0.31 mm, respectively.
For the models with $a_{\rm max}=1.05$, 2.92 and 10.72 mm, we can obtain only the lower limit of the disk mass as 79, 90 and 132, respectively, since the disk is optically thick even at ALMA Band 4.

From these estimates, we conclude that the disk mass would be $\sim$ 250$M_{\oplus}$ if the maximum dust size is $\sim140~{\rm \mu m}$ as inferred from the Band 7 polarimetric observation and at least $\sim80M_{\oplus}$ even if the dust size is larger than $\sim140~{\rm \mu m}$.
In any case, the CW Tau disk is quite massive and has an enough amount of dust to form cores of giant planets, even though the disk size is relatively small.

\subsection{Surface density transition at 50 au}
Our results clearly shows a very steep cutoff of the dust surface density at $\sim$ 50 au.
One can interpret this steep cutoff as an outer edge of the dust disk (e.g., \citealt{BA14}), since
most of the disk emission and mass are contained within it.
However, our models show that the dust disk still extend beyond 50 au, with the surface density reduced by a factor of $\sim$ 10 at 50 au.
This unique structure may be related to the so-called dead-zone outer edge (e.g., \citealt{Dzyurkevich+13,Turner+14}).
At the dead-zone outer edge, the turbulent viscosity arising from the magneto-rotational instability increases from inside out.
This results in a steep decrease in the gas surface density (e.g., \citealt{Delage+21}), which can be also seen in the dust distribution (e.g., \citealt{Kretke+09,Ueda+21c}).
If this is the case, the similar transition in the gas surface density as well as the transition in the turbulence strength should be observed.
Therefore, to characterize this unique structure in more detail, observations with molecular line emissions which trace the gas mass and kinematics are needed.

The other possibility is that the outer region contains unresolved ring-like substructures, as high resolution ALMA observations have revealed for many disks (e.g., \citealt{Isella+16,Fedele+18,Kudo+18,Perez+19}).
If unresolved ring-like structures are present, such structures may be seen as a faint smooth disk.
As mentioned in Section \ref{sec:diskmodelling}, the signal-to-noise ratio is $\sim$ 2--3 in the outer region and hence it is difficult to quantify the outer disk properties robustly.
Higher sensitivity observations are necessary to construct a detailed model for the outer region of the CW Tau disk.

\subsection{Origin of the dust gap} \label{sec:gap}
Gap structure in dust disks can be created by a decrease in either dust optical depth (i.e., dust density and/or dust opacity; e.g., \citealt{Pinilla+12,Dipierro+15}) or temperature (e.g., \citealt{Ueda+21}).
If the gap is created by the temperature variation, it can be seen even in the optically thick regime \citep{Ueda+21}.
However, the gap in the CW Tau disk is not detected in our Band 8 image, suggesting that the gap structure is induced by the dust optical depth.
The radial intensity profile at ALMA Band 4 shows the gap width of $\sim$ 4 au, which is much smaller than the spatial resolution of the data.
Therefore, higher resolution observation is necessary to constrain details of the gap and its origin.

\subsection{The CW Tau disk in the context of survey studies}
As shown in this work, the CW Tau disk has many interesting characteristics.
In this section, we compare the physical quantities which have been often used for characterizing disks with those of the CW Tau disk.

\begin{figure}[ht]
\begin{center}
\includegraphics[scale=0.46]{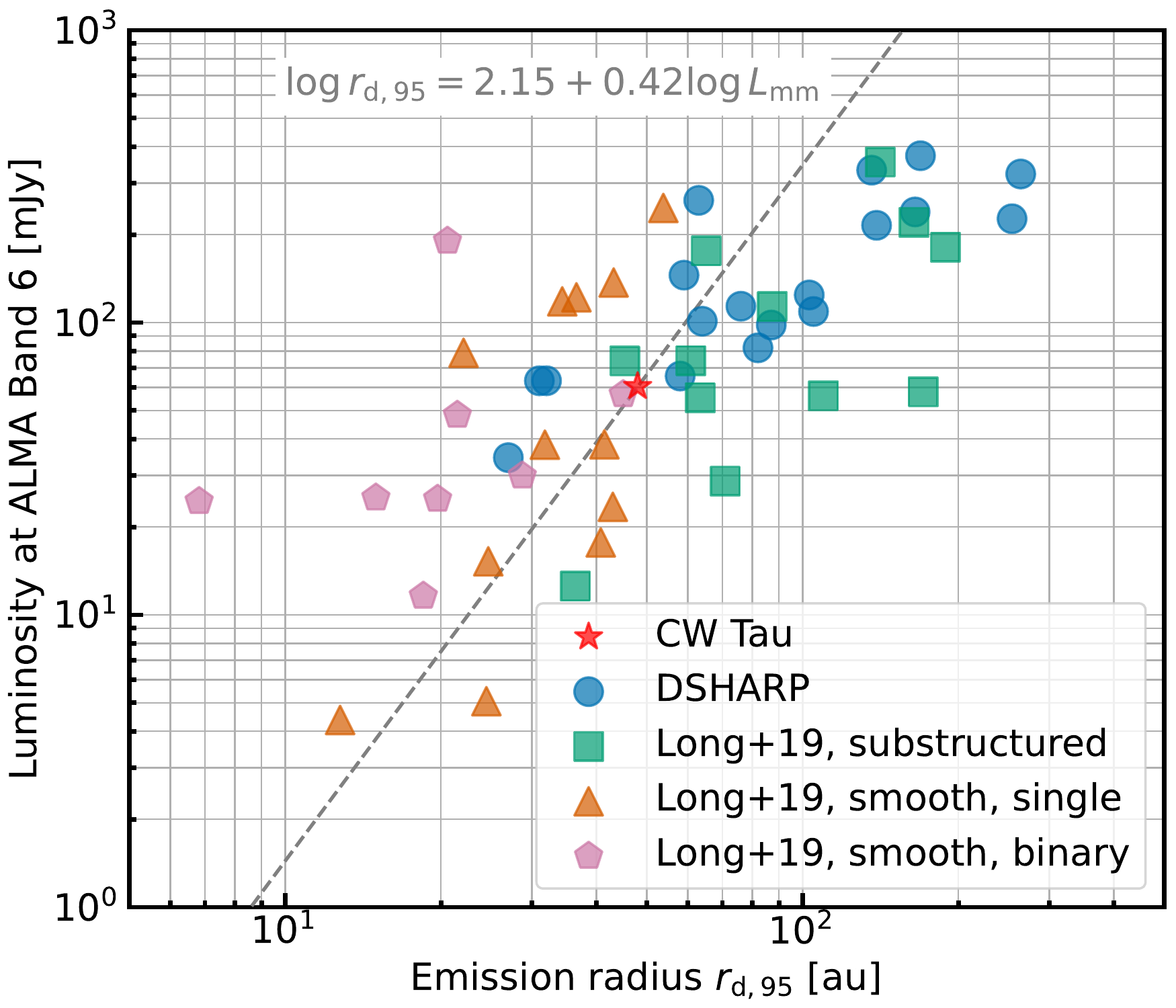}
\caption{
Comparison of the size-luminosity relation of the CW Tau disk with previously studied disks \citep{Andrews+18,Long+19}.
The luminsity is scaled to a distance of 140 pc.
The gray dashed line shows the empirical relation of the mm luminosity and radius derived by \citet{Long+19}.
}
\label{fig:L-R}
\end{center}
\end{figure}

Figure \ref{fig:L-R} compares the relation between mm luminosity $L_{\rm mm}$ and emission radius $r_{\rm d,95}$ of the CW Tau disk measured at ALMA Band 6 with disks observed by previous studies \citep{Andrews+18,Long+19}.
The disk luminosity is known to be roughly proportional to the square of the disk size at mm wavelengths \citep{Tripathi+17,Barenfeld+17,Andrews+18a,Long+19}.
The less-biased survey of disks in Taurus region has shown that the majority of Taurus disks is more compact than DSHARP disks \citep{Long+19}.
As clearly seen in Figure \ref{fig:L-R}, the CW Tau disk is compact compared to the DSHARP disks and lies between large substructured population and compact smooth (no substructures are found with $0\farcs12$ resolution) population.
The size-luminosity relation of the CW Tau disk follows the relation derived by the less-biased survey (\citealt{Long+19}).
This means that, even though the CW Tau disk has interesting characteristics (very massive and compact), the CW Tau disk is {\it normal} in the size-luminosity diagram compared to the other disks.

The CW Tau disk is {\it normal} even in terms of the spectral index at mm wavelengths that has been used for characterizing the dust size and mass in disks.
Although the CW Tau disk shows a steep spectral slope of $\sim3.7$ between 2.17 and 3.56 mm, it is flatter between 0.89 and 3.56 mm, $\alpha\sim2.7$.
This is because the CW Tau disk is optically thick at 0.89 mm.
As reported by \citet{Tazzari+21}, the spectral slope between 0.9 and 3 mm is typically $\lesssim$ 2.5 in Class II disks, which is interpreted as the disk is optically thick or dust grains are large.
Although our spectral slope between 0.89 and 3.56 mm is comparable to that derived by \citet{Tazzari+21}, the slope between 2.17 and 3.56 mm is much steeper and more consistent with that of small grains.
This indicates that the spectral slope derived from ALMA Band 7 and 3 might be significantly affected by the effect of optical thickness.
If the disk is optically thick, we may overestimate the dust size (and hence the mm opacity) from the spectral index, which is crucial for disk mass estimates.
To accurately estimate the dust size, we need two wavelengths at which the disk is optically thin.
As ALMA Band 3 and 7 (or 6) are often used to estimate the spectral index, we propose to use Band 4 instead of Band 7 for more precise estimate of dust size.

\section{Summary}\label{sec:conclusion}
We reported the high resolution ALMA multi-band observations toward a compact protoplanetary disk around CW Tau.
Our key findings are as follows.
\begin{itemize}

\item The image of the CW Tau disk at ALMA Band 4 clearly shows a dust gap at $\sim$ 20 au, although it is expected to be not spatially resolved with the current resolution ($\sim0\farcs08$).

\item The SED between ALMA Band 6 and 8 follows the spectral slope of $2.0\pm0.24$, indicating that the disk is optically thick, while the slope is $3.7\pm0.29$ between Band 4 and 3.56 mm, implying that the dust size is small ($\lesssim$ 1 mm).

\item The parametric fitting of the radial intensity profiles shows that the combination of estimated temperature profile and gas surface density (100 times of dust) indicates the Toomre's Q value is $\sim$ 1--3 at most of the disk outer region.

\item Combining our ALMA data with the previous 3.56 mm observation, the typical dust size in the CW Tau disk is preferred to be $\lesssim 1$ mm with 1-$\sigma$ confidence, but larger grains ($\gtrsim 1$ mm) are still acceptable with 2-$\sigma$ confidence.

\item The total dust mass is estimated as $\sim$ 250$M_{\oplus}$ for the maximum dust size of $\sim140~{\rm \mu m}$ that is inferred from the previous Band 7 polarimetric observation and at least $80M_{\oplus}$ even for larger grain sizes.

\item The CW Tau disk is {\it normal} in terms of the disk luminosity, radius and spectral index at ALMA Band 6, although it is quite massive compared to the typical disk mass reported by the survey studies.

\end{itemize}

These interesting characteristics of the CW Tau disk hint the importance of the multi-band analysis toward the compact disks.

\acknowledgments
We thank an anonymous referee for comments that significantly improve this manuscript.
We also thank Kiyoaki Doi for useful comments.
This work was supported by JSPS KAKENHI Grant Numbers 18K13590, 19J01929 and 20K04017.
This paper makes use of the following ALMA data: ADS/JAO.ALMA$\#2019.1.01108.{\rm S}$.
ALMA is a partnership of ESO (representing its member states), NSF (USA) and NINS (Japan), together with NRC (Canada), MOST and ASIAA (Taiwan), and KASI (Republic of Korea), in cooperation with the Republic of Chile. The Joint ALMA Observatory is operated by ESO, AUI/NRAO and NAOJ. 



\bibliographystyle{aasjournal}
\bibliography{reference}

\end{document}